\documentstyle[multicol,aps,epsf]{revtex}
\def\d{{\rm d}}
\begin{document}
\title{Motion-Induced Radiation from a Dynamically Deforming Mirror}
\author{Faez Miri and Ramin Golestanian}
\address{Institute for Advanced Studies in Basic Sciences,
Zanjan   45195-159, Iran}
\date{\today}
\maketitle
\begin{abstract}
A path integral formulation is developed to study the spectrum of radiation from a 
perfectly reflecting (conducting) surface. It allows us to study arbitrary 
deformations 
in {\it space} and {\it time}. The spectrum is calculated to second order in the
height function. For a harmonic traveling wave on the surface, we find many 
different regimes in which the radiation is restricted to certain directions. 
It is shown that high frequency photons are emitted in a beam with
relatively low angular dispersion whose direction can be controlled by the 
mechanical deformations of the plate. 
\end{abstract}
\pacs{42.50.Lc, 42.50.Dv, 03.65.-w, 12.20.Ds}
\begin{multicols}{2}

The Casimir effect \cite{Casimir} provides a direct link between
the macroscopic world and the quantum vacuum.  
Imposing boundary conditions on the 
electromagnetic field in the space between two parallel conducting 
plates changes zero point vacuum fluctuations, and results
in an attractive force between the plates. Dynamical modification 
of boundary conditions
tends to perturb the quantum vacuum, leading to excitation
of photons\cite{Moore}. The corresponding {\it radiation reaction }
results in a friction-like dissipative force. This is one of the manifestations
of the so-called dynamic Casimir effect
\cite{Fulling,Jaekel,Cavity,GK}. Dissipative forces
also appear for nonideal conductors that move laterally \cite{Ohmic}. 
However, they are fundamentally different, because in this case the
dissipation mechanism is due to the Ohmic loss of the induced current
in the bulk.

The emission of photons by a perfect cavity, and the observability
of this energy, has been studied by different approaches
\cite{Cavity,Netodouble}. The most promising candidate is
the  resonant production of photons when the mirrors
vibrate at the optical resonance frequency of the cavity\cite{Davis}. 
The radiation due to vacuum fluctuations of a 
collapsing dielectric sphere (bubble) has been studied and proposed 
as a possible explanation for the intriguing phenomenon of sonoluminescense
\cite{Eberlein}.
Recently Maia Neto and Machado \cite{Netoradiation} 
studied the angular distribution and frequency spectrum of radiation
from a single perfectly reflecting mirror with oscillatory motion. 
They found restrictions on the angular distribution of radiation, i.e. not 
all directions are allowed for the radiation. 

The above calculations are for rigidly oscillating flat plates and spheres.
In this paper, we have developed a path integral method to study the radiation
from perfectly reflecting mirrors that undergo small dynamic deformations.
We calculate the angular distribution and the total spectrum of radiation from
a dynamically perturbed quantum vacuum. We examine the specific example
of a single perfectly reflecting plate undulating harmonically with a frequency
$\omega_0$ and a wavevector ${\bf k}_0$. 
We find that depending on the value
of $\omega_0/k_0$,  radiation at a frequency $\Omega$ can belong to 
different  classes, each charachterized by specific constraints on the angular 
distribution that restrict the radiation to a particular solid angle window. 
Radiation at frequencies close to $\omega_0$ are
restricted to a single direction which is parallel to ${\bf k}_0$ and makes an 
angle $\theta_b=\sin^{-1}(k_0/\omega_0)$ with the surface normal. Hence,
we can control the direction of emission as well as the frequency and 
the angular dispersion 
of the beam just by tuning  $(\omega_0,{\bf k}_0)$.

The total spectrum of radiation is found by integrating the angular
distribution of radiation over the unit sphere. It is found to be a symmetric 
function
with respect to $\omega_0/2$, where it is peaked. The peak sharpens as the
parameter $\omega_0/k_0$ is increased, and saturates for $k_0=0$.
The connection between the dissipative dynamic Casimir force and
radiation of photons is made explicit by calculating the total energy of
photons radiated per unit time per unit area of the plate, which is
identical to the energy dissipation rate calculated in Ref.\cite{GK}.
We find that no radiation exists at frequencies higher than $\omega_0$
, due to conservation of energy, and also for $\omega_0/k_0<1$ in agreement with 
Ref.\cite{GK} who find no dissipative forces for this regime.

The calculation is comprised of three steps. The LSZ formalism is used
to relate the transition amplitude from an empty vacuum (at $t \to -\infty$)
to a state with two photons  (at $t \to +\infty$), to two-point correlation
functions of the field.  The two-point correlations can then be calculated 
perturbatively in the deformations using a path integral formulation. Finally,
we obtain the radiation spectra by integrating over the state of one photon.

Our approach is a natural generalization of the path integral method developed
by Golestanian and Kardar \cite{GK} to study mechanical response of vacuum.
As in Ref.\cite{GK}, we simplify the problem by considering
a scalar field $\Phi$ (in place of the electromagnetic vector
potential) with the action $(c=1)$
\begin{equation}
S=\frac{1}{2} \int \d^d X \;\partial_{\mu}\Phi(X)
			\partial_{\mu}\Phi(X),\label{action}
\end{equation}
where summation over $\mu=1,\cdots,d$ is implicit. 
Following a Wick rotation, imaginary time appears as another
coordinate $X_d=i t$ in the $d$--dimensional space-time. 
We would like to quantize the field subject to the constraints 
of its vanishing on a set of $n$ manifolds (objects) defined
by $X=X_{\alpha}(y_{\alpha})$, where $y_{\alpha}$ parametrize 
the $\alpha$th manifold. Following Refs.\cite{GK,LiK}, we can impose 
the constraints using delta functions, and calculate the two-point correlation 
function of the field as $(\hbar=1)$
\begin{eqnarray}	\label{G1}
G(X,Y)&=&\int {\cal D}\Phi(X) \;\Phi(X) \Phi(Y) \\
	&& \times \;	\prod_{\alpha=1}^{n} \prod_{y_{\alpha}}
		\delta\left(\Phi\left(X_{\alpha}(y_{\alpha})
		\right)\right)\;{\rm e}^{-S[\Phi]}.\nonumber
\end{eqnarray}
Now  consider a scattering process in which we start with a vacuum as the initial 
state,
perturb it with dynamic boundary conditions during a finite period of time, and 
look for particles
(that have possibly been created as a consequence of the perturbation) in the 
final state of the system. Using the LSZ reduction formalism \cite{Greiner} ,  we 
can find the transition amplitude from the vacuum to a two photon state from the 
matrix element
s of the $S$-matrix, as \cite{twophoton}
\begin{eqnarray} \label{Sfi1}
 S_{fi}&\equiv& \left< k,k' \mid {\hat S} \mid 0 \right> \\
&=&\frac{1}{2  \sqrt{\Omega_k {\Omega'}_{k'}} }
\tilde{G}(k,\Omega_k;k',\Omega'_{k'})  (k^2-\Omega_k^2)({k'}^2-{\Omega'}_{k'}^2), 
\nonumber
\end{eqnarray}
where $k=({\bf k}, k_z)$, and $\tilde{G}(k,\Omega;k',\Omega')$ is the Fourier 
transform 
of the two point function defined in Eq.(\ref{G1}), and $\Omega_k=\sqrt{{\bf 
k}^2+k_z^2}$ 
denotes the photon dispersion relation in vacuum \cite{pole}.
To obtain the distribution of radiation as measured  by a detector, we need to 
sum over the momenta of the second photon. The angular distribution of  
radiated photons is given by 
$
{\cal P}(\Omega, \theta,\phi)= (\Omega^2/8 \pi^3) \int \d^3 k'/(2\pi)^3 
\; \mid S_{fi} \mid^2,
$
and the total number of photons by
$
{\cal P}(\Omega)=\int \d \cos\theta \; \d \phi   \; {\cal P}(\Omega, \theta,\phi),
$
where $\theta$ and $\phi$ are the polar angles defining the direction of the 
emission
of radiation. 

The calculation of the two-point function in Eq.(\ref{G1}) is generally 
complicated
since the Lagrange multiplier fields are defined on a set of manifolds
with nontrivial geometry. To be specific, we focus on  a 2d plate
embedded in 3+1 space-time. Deformation of the plate is parametrized 
by the height function $h({\bf x},t)$ along the $x_3$-axis, where 
${\bf x}\equiv(x_1,x_2)$ denotes the two lateral 
space coordinates while $t$ is the time variable. Generalizing the
method of Ref.\cite{GK}, we can calculate $G(X,Y)$ by a perturbative series in
powers of the height function. The resulting expression for the 
Fourier transform of  time ordered two point function reads
\begin{eqnarray} \label{GKWKW}
\tilde{G}(k,\Omega;k',\Omega')&=&\frac{2k_z \sqrt{{\Omega'}^2- {{\bf k}'}^2}
+2k'_z\sqrt{\Omega^2- {\bf k}^2}}
{({\bf k}^2+k_z^2-\Omega^2)({{\bf k}'}^2+{k'_z}^2
-{\Omega'}^2)}  \\
&&\;\;\times \; \tilde{h}( {\bf k}+{\bf k}', \Omega+\Omega') 
+O(h^2) , \nonumber
\end{eqnarray}
to the leading order. The corresponding expression for the 
spectrum reads
\begin{equation}  	\label{PRh}
{\cal P}(\Omega,\theta,\phi)=\int \frac{\d \omega \d^2 {\bf q}}{(2\pi)^3} \;
{\cal R}(\Omega,\theta,\phi; {\bf q},\omega) \;  \mid \tilde{h}({\bf 
q},\omega)\mid^2,
\end{equation} 
in which
\begin{eqnarray}	\label{Rdef}
{\cal R}(\Omega,\theta,\phi; {\bf q},\omega) &=&
 \frac{\Omega}{2\pi^3}  (\Omega^2-{\bf k}^2)
\sqrt{(\omega-\Omega)^2-({\bf q}-{\bf k})^2}  \\ 
&& \times \; \Theta (\omega-\Omega)  \;
\Theta  \left( (\omega-\Omega)^2-({\bf q}-{\bf k})^2 \right) ,\nonumber 
\end{eqnarray}
where ${\bf k}=(\Omega \sin\theta \cos \phi,\Omega \sin\theta \sin \phi) $
, and $\Theta(x)$ is the Heaviside step function.

As a concrete example, let's consider a case where the deformation of
the plate is in the form of a harmonic traveling wave. The height function 
can be represented as
$ h({\bf x},t)=d \cos({\bf k}_0 \cdot {\bf x}-\omega_0 t) $
for a harmonic wave that propagates with the phase velocity 
$\omega_0/k_0$. This helps us study the optical reponse of vacuum
to surface deformations in the frequency--wavevector domain.
Most of the following analysis deals with the far from trivial
problem of matching the spatial and temporal sinusoids characterizing
the mirror to those characterizing the two plane-wave photons.
This problem, which is essentially kinematic, has strong echoes of
more familiar though somewhat simpler problems involving crystal
and other gratings.

The angular distribution of radiated photons per unit time per unit area
of the plate $P$, can be calculated from Eq.(\ref{Rdef}) as
$(P \equiv{\cal P}/A T )$
\begin{eqnarray}	\label{Pangular2}
P(\Omega,\theta,\phi)
 &=& \frac{d^2 \Omega}{4\pi^3}  (\Omega^2-{\bf k}^2)
\sqrt{(\omega_0-\Omega)^2-({\bf k}_0-{\bf k})^2}  \\ 
&& \times \; \Theta (\omega_0-\Omega)  \; 
\Theta  \left( (\omega_0-\Omega)^2-({\bf k}_0-{\bf k})^2 \right) .\nonumber 
\end{eqnarray}
The first step function in the above expression shows that the
emitted photons can only have frequencies smaller than $\omega_0$.
(The negative frequency Fourier component of the height function drops
out due to similar constraints.) 
The second step function places some restrictions on the possible directions
for emission of radiation, leading to a nontrivial angular distribution. If we 
take
${\bf k}_0$ to be parallel to the $x_1$ direction, the condition on possible
directions for the radiation reads
\begin{equation} 	\label{condition1}
 (\omega_0-\Omega)^2 > k_0^2+\Omega^2 \sin^2 \theta
- 2 k_0 \Omega \sin \theta \cos \phi.
\end{equation}

The above inequality can be most easily solved geometrically.
Using the parametrization 
$(x,y,z)=(\sin\theta \cos\phi,\sin\theta \sin\phi, \cos\theta)$ for the unit
director, the permissible directions for radiation can be represented
as regions of the $xyz$-space that satisfy the following conditions:
$[(\omega_0-\Omega)/\Omega]^2 >(x-k_0/\Omega)^2+y^2 $,
which corresponds to the interior of a cylinder with radius
$r=(\omega_0-\Omega)/\Omega$,  and $x^2+y^2+z^2=1$ which
corresponds to surface of a unit sphere.  Hence, we simply need to
find the intersection of the unit sphere and the cylinder.
Depending on the radius of the cylinder and on the way they
intersect many different cases may arise, each of which leading to
a characteristic angular distribution of radiation. The partition of the
parameter space into these different regions is summarized in Fig.~1.

In the special case $k_0=0$, there are two possibilities. For $r>1$,
the sphere is completely contained within the cylinder, and there
is no restriction for radiation. For $r<1$, however, the intersection
restricts the radiation zone to the circular patch defined by
$\theta < \theta_0=\sin^{-1}((\omega_0-\Omega)/\Omega)$ \cite{Netoradiation}.
In general $k_0$ is not zero, and there exist 7 interesting regimes:
{\bf (1)} For $k_0/\Omega<1 $, $ r <1-k_0/\Omega$,  and $r <  k_0/\Omega$,
the projection of the cylinder onto the $xy$-plane is completely contained
within the unit circle, but it does not contain the origin. In this case
all radiation is limited to $ -\phi_{max} <\phi <\phi_{max} $,
and (for a specific angle $\phi$) $\theta_{+}(\phi) < \theta < \theta_{-}(\phi)$ 
where $\phi_{max} =\sin^{-1}[(\omega_0-\Omega) k_0/\Omega^2] <\pi/2$,
and
\begin{eqnarray}	\label{theta+-}
\theta_{\pm}(\phi)&=&\cos^{-1}
\left\{1-\left[(\omega_0-\Omega)/\Omega\right]^2
-(k_0/\Omega)^2 \cos 2\phi \right.\\
&&\left. \pm 2 (k_0/\Omega)^2 \cos\phi 
\left[ (\omega_0-\Omega)^2/k_0^2-\sin^2\phi \right]^{1/2} \right\}^{1/2}. 
\nonumber
\end{eqnarray}
It is interesting to note that in this regime no radiation is emitted normal
to the plate. In fact for $\Omega \sim \omega_0 ~ (r \ll 1)$, the radiation
is limited to $\theta \approx \sin^{-1}(k_0/\omega_0)$, and $\phi \approx 0$.
Furthermore, the beam will have a relatively
low angular dispersion ( $\sim r$).
{\bf (2)} In the regime $k_0/\Omega<1 $, and $1-k_0/\Omega  <  r < k_0/\Omega$,
the projection of the cylinder onto the $xy$-plane intersects with the unit
circle with its center being inside, but it does not contain the origin.
In this case there is a critical angle defined as
\begin{equation}	\label{phic}
\phi_c=\cos^{-1}\left(\frac{1-\left[(\omega_0-\Omega)/\Omega\right]^2
+(k_0/\Omega)^2}{2k_0/\Omega}\right),
\end{equation}
at which the behavior changes drastically. For  $-\phi_c <\phi<\phi_c$, we have  
$\theta_{+}(\phi)<\theta < \pi/2$, while for  $ \phi_c <\phi<\phi_{max}$ or 
$ -\phi_{max} <\phi<-\phi_c$, we have  $\theta_{+}(\phi)<\theta<\theta_{-}(\phi)$.
{\bf (3)} When $r > 1- k_0/\Omega$, 
and $k_0/\Omega  <  r < k_0/\Omega+1 $, the projection of the cylinder onto 
the $xy$-plane intersects with the unit circle, and contains the origin.
In this case for $-\phi_c <\phi<\phi_c$, we have $0<\theta <\pi/2$, and for
$ \phi_c< \phi<\pi  $  or $ -\pi <\phi <-\phi_c $, we have
 $0<\theta<\theta_{-}(\phi)$.
{\bf (4)} In case $ r >1+k_0/\Omega$,
the projection of the cylinder onto the $xy$-plane completely contains
the unit circle. Radiation can be detected in all dirctions with no
restrictions, i.e. $0<\theta <\pi/2$ and $0 <\phi <2\pi $.
{\bf (5)} For $k_0/\Omega >1 $, and  $r < k_0/\Omega -1 $,
the projection of the cylinder onto the $xy$-plane is disjoint from
the unit circle. No radiation exists in this case.
{\bf (6)} When
$k_0/\Omega >1 $, and  $k_0/\Omega-1 <r < k_0/\Omega$,
the projection of the cylinder onto the $xy$-plane intersects with the unit
circle with its center being outside, but it does not contain the origin.
Radiation is limited to $-\phi_c <\phi<\phi_c$ and $\theta_{-}(\phi)<\theta 
<\pi/2$.
{\bf (7)} Finally, for $k_0/\Omega <1 $, and $k_0/\Omega< r < 1-k_0/\Omega$,
the projection of the cylinder onto the $xy$-plane is completely contained
within the unit circle, and it contains the origin. In this case
we have $0<\theta <\theta_{-}(\phi)$  for $0 <\phi<2\pi$.

The dashed lines in Fig.~1 represent three different classes of trajectories 
that are characterized by the parameter $\omega_0/k_0$. In each class,
radiation has peculiar angular distributions. The
total number of photons radiated per unit time per unit area, $P(\Omega)$, is 
plotted for these three different classes in Fig.~2. The curves are obtained
by numerical integration of the angular distribution
taking into account the complicated boundary conditions discussed above.
The plots are symmetric with respect to $\Omega/\omega_0=1/2$, where they
are peaked. The symmetry is a manifestation of the process being
due to a two-photon emission. As  $\omega_0/k_0$ is increased, the peak
becomes narrower and sharper until it saturates at  $\omega_0/k_0=\infty$
\cite{Analytical}.

We can calculate the total radiated energy of photons per unit time per unit area
to obtain the energy dissipation rate. Given the symmetry of the spectrum 
(the average frequency of the radiation is always $\omega_0/2$), we obtain
$R=\int \d \Omega  \Omega P(\Omega)=(\omega_0/2) 
\int \d \Omega  P(\Omega)$. The energy dissipation rate can also be calculated
from the mechanical reponse of the system \cite{GK}, and it reads
$R=(d^2 \omega_0/720 \pi^2)(\omega_0^2-k_0^2)^{5/2}$.
Conservation of energy requires the two methods to give the same result
for the total number of radiated photons, namely
it requires
\begin{equation}
{\cal N}=\int \d \Omega  \; {\cal P}(\Omega)=\frac{d^2 c T A}{360 \pi^2} 
\;(\omega_0^2/c^2-k_0^2)^{5/2},
\end{equation}
where the factors of $c$ are restored to allow estimation in physical units. 
The above identity was found to be true for each class,
within the accuracy of our numerical integrations.

In conclusion, we have developed a path integral formulation to study
the radiation from perfectly reflecting mirrors with dynamic fluctuations.
Although its application to a simplest example lead to
relatively rich and complicated behaviors, the method itself is quite general.
Application of the method to various other geometries, such as a fluctuating wire 
in 3+1D space-time, two fluctuating plates, and arrays of such objects
will be published elsewhere.

It is a pleasure to acknowledge M. Kardar, V. Karimipour, M.R.H. Khajehpour,
M. Khorrami, and A. Shariati for invaluable discussions and suggestions. 
We also wish to thank M. Kardar for a careful reading of the manuscript and
giving critical comments. RG acknowledges supprot 
from NSF grant DMR-93-03667 during a visit to MIT.

\begin{figure}
\epsfysize=4.0truein
\centerline{\epsffile{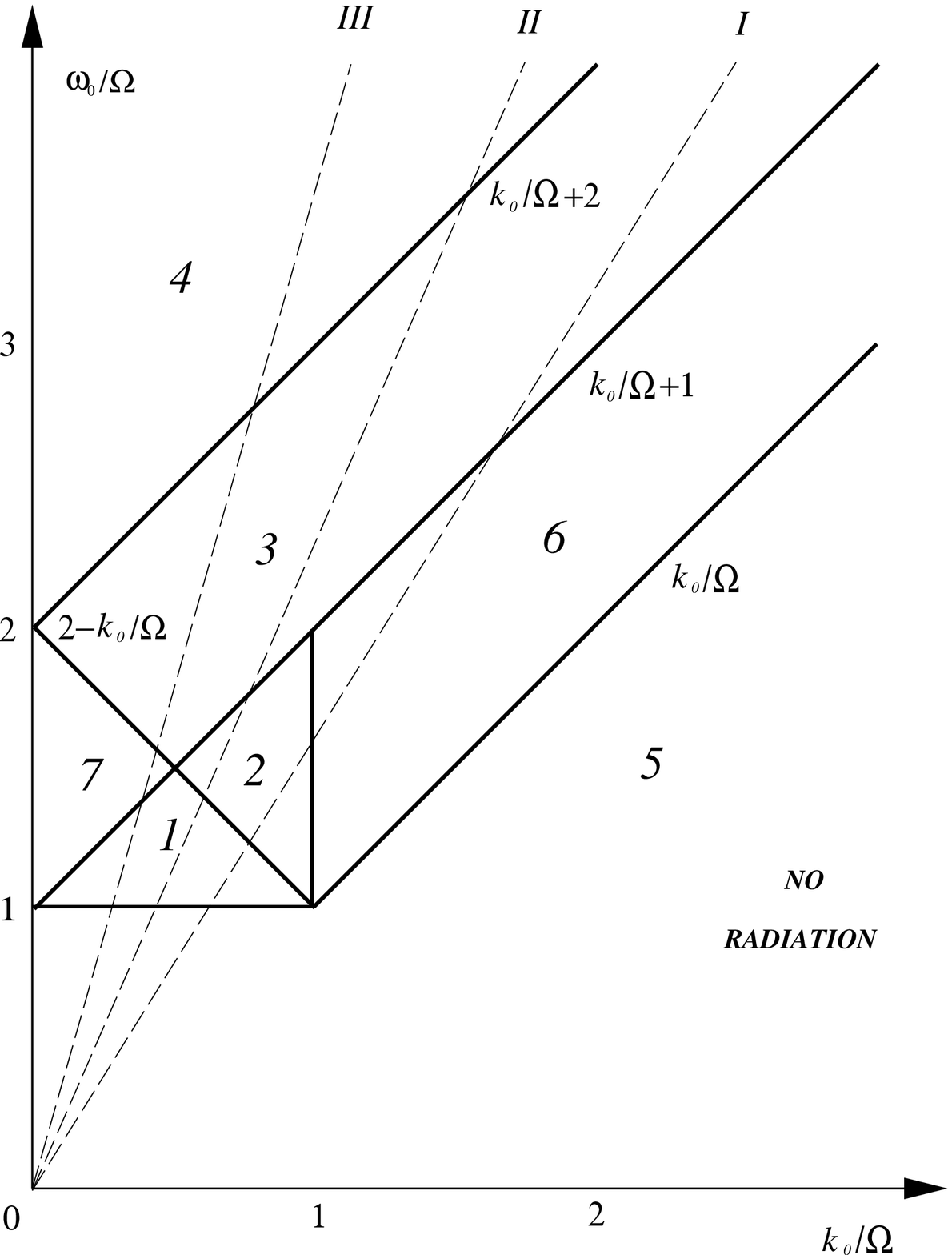}}
FIG.~1. Different regions in the frequency--wavevector plane.
There is no radiation for 
$0<\omega_0/k_0<1$. Class $I$ corresponds to $1<\omega_0/k_0<2$,
class $II$ corresponds to $2<\omega_0/k_0<3$, and 
class $III$ corresponds to $3<\omega_0/k_0< \infty$. 
\label{Fig1}
\end{figure}

\begin{figure}
\epsfysize=2.1truein
\centerline{\epsffile{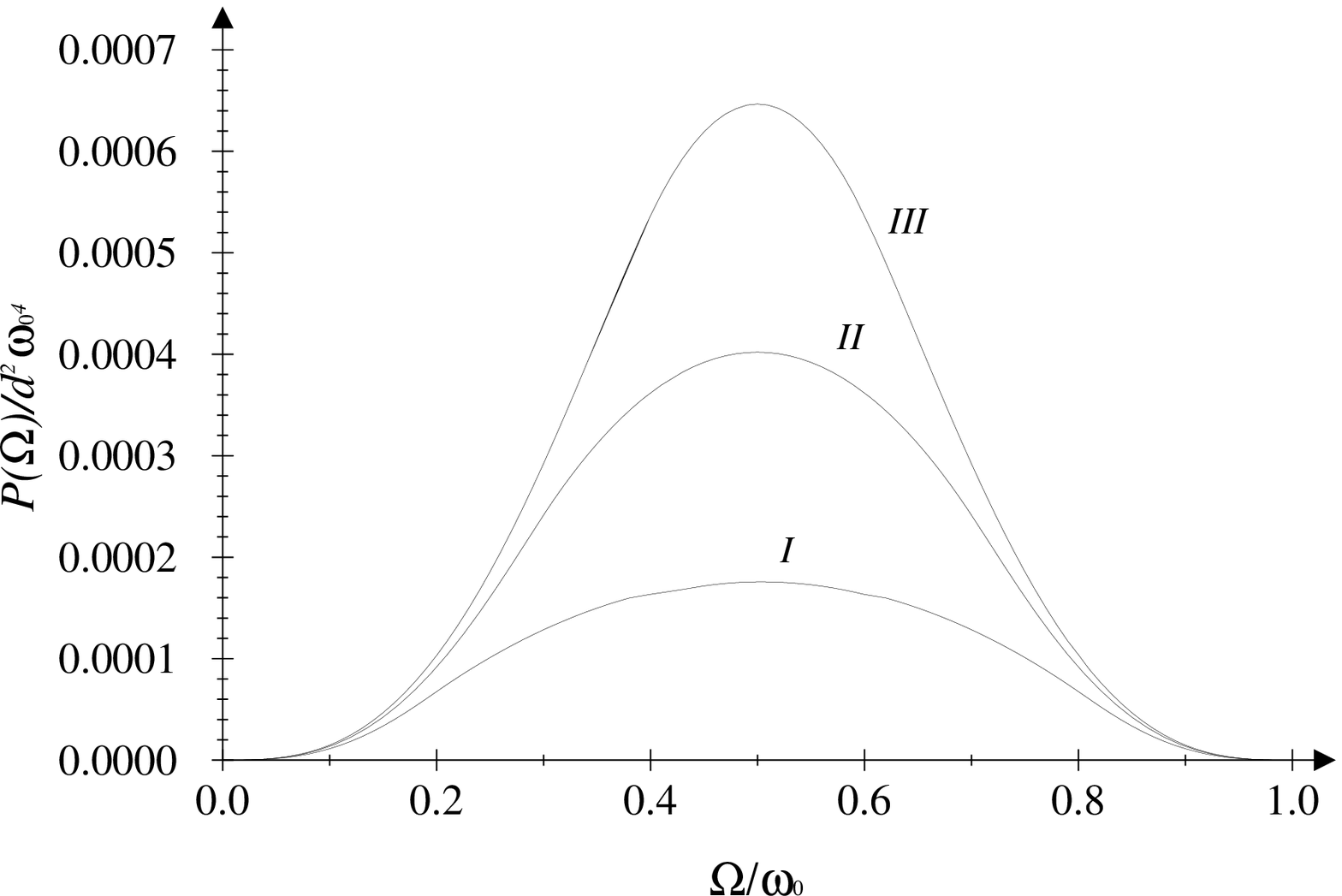}}
FIG.~2. Dimensionless spectrum of radiation 
$P(\Omega)/(d^2 \omega_0^4)$ for different classes.
Plot $I$ corresponds to $\omega_0/k_0=5/3$,
plot $II$ corresponds to $\omega_0/k_0=5/2$, and
plot $III$ corresponds to $\omega_0/k_0=5$.
\label{Fig2}
\end{figure}

\end{multicols}


\begin{references}
\bibitem{pole}
Note that the dispersion relation sets $ (k^2-\Omega_k^2) $ to 
zero, so we only get nonzero matrix elements in Eq.(\ref{Sfi1}) 
if we have the same poles in the two point function.
\bibitem{Analytical}
In this case the analytical expression for the spectrum can be worked out as
$P(\Omega)= (d^2  \omega_0^4/{128\pi^2})  
[1-(2\Omega/\omega_0-1)^4 
+(2\Omega/\omega_0-1)^2 \ln |2\Omega/\omega_0-1|].$
This expression is identical (within a factor of 2) to that obtained in 
Ref. \cite{Netoradiation}
for the TE mode of the electromagnetic (EM) field for $\omega_0/k_0=\infty$, and
has no divergence at $\Omega/\omega_0=1/2$ . Interestingly, the TM mode 
of the EM field is shown in Ref.\cite{Netoradiation} to have a logarithmic 
divergence at  $\Omega/\omega_0=1/2$ for  $\omega_0/k_0=\infty$, which we 
anticipate to be 
rounded off for any finite $\omega_0/k_0$.
\end{references}
\end{document}